# Characteristics of shock tube generated compressible vortex rings at very high shock Mach numbers

Sajag Poudel, Chandrala Lakshmana Dora, Ashoke De, Debopam Das*

Department of Aerospace Engineering, Indian Institute of Technology Kanpur, Kanpur, 208016, India.

*Corresponding Author (das@iitk.ac.in)

**Abstract**

Compressible vortex rings, usually formed at the open end of a shock tube, often show interesting phenomena during their formation, evolution, and propagation depending on the shock Mach number ($M_s$) and exit flow conditions. The Mach number of the translating compressible vortex rings ($M_v$) investigated so far in the literature is subsonic as, the shock tube pressure ratio (PR) considered is relatively low. In this numerical study we focus on low to high vortex ring Mach numbers ($0.31 < M_v < 1.08$) cases with a particular focus on very high $M_v$ cases that are not been reported in experiments as, it is difficult to obtain in laboratory. Using hydrogen as a driver section gas inside the shock tube, a supersonic compressible vortex ring ($M_v > 1$) is obtained for first time.

It is established that the SST k-ω based DES turbulent model replicates the experimental observation better than the previously published results at different stages of development of the vortex ring. DES, which is an inbuilt hybrid of LES and RANS approaches is evoked that can automatically switch to the sub-grid scale (SGS) model in the LES regions (i.e. with different scale vortical structures) and to a RANS model in the rest of the region (i.e. where the grid spacing is greater than the turbulent length scale). The DES model can predict characteristics of the shear layer vortices as well as counter-rotating vortex rings (CRVRs) as reported in the experimental measurements.

Formation of multiple triple points and the corresponding slip-stream shear layers and thus multiple CRVRs behind the primary vortex ring at different radial locations, in addition to the usual CRVRs, appears to be a unique characteristic for high Mach number vortex rings. For high PR, $H_2$, case during formation stage, a vortex layer of reverse circulation (that of primary vortex ring) is formed from the outer wall of the shock tube, whose instability formed another series of opposite circulation vortices which interfere with the primary vortex ring considerably. Also, near the central zone, a near stationary slipstream vortex and multiple, fast moving tiny vortices of opposite circulation to slipstream vortex are observed. Mechanisms for formation of these complex vortex structures are identified. The implications of these phenomena on vortex rings' geometric and kinematic characteristics such as ring diameter, core diameter, circulation, translational velocity as well as vortex Mach number are discussed in detail illustrating their differences with low PR cases.

**1. Introduction**

Vortex ring is a toroidal vertical structure in a fluid with a rotational core and irrotational outer fluid. Vortex rings are observed in many unsteady flow configurations such as in, artillery gun firing, volcanic eruptions, starting flow from a rocket nozzle, left ventricle of a human heart during cardiac diastole and many other engineering and physical problems. In general, a vortex ring is produced when a slug of fluid is ejected impulsively from a nozzle or an orifice. Although, several studies exist in the literature for incompressible vortex rings (Maxworthy 1974, Gharib, Rambod & Shariff 1988, James & Madnia 1996, Danaila and Hélie 2008, Moore & Pullin 1998, Elder & De Hass 1952, Sullivan et al. 2008, Das, Bansal & Manghnani 2017), number of studies available for compressible vortex rings is limited (Moore & Pullin 1998, Elder & De Hass 1952, Baird 1987, Dora et al. 2014). Typically, shock tube with an open end is used for producing an impulsive high-speed flow that generates a compressible vortex ring at the sharp exit. The rapid variations in velocity, density, and vorticity across the vortex ring core make it quickly turbulent at this high Reynolds numbers. The flow field of compressible vortex ring is complicated due to the diffraction of incident shock at sharp exit, expansion of flow, the formation of embedded shock, Mack disc, and slipstream shear layers at the triple point, particularly at high Mach numbers (Moore & Pullin 1998, Elder & De Hass 1952, Baird 1987, Dora et al. 2014).

The formation of a vortex ring at the open end of a shock tube can be described in two stages. First, the shock diffraction and the initial axisymmetric roll-up of the shear layer (of the shock tube wall) separated at nozzle lip and second, the subsequent evolution and propagation of the vortex ring (Elder & De Hass 1952, Baird 1987, Dora et al. 2014). Depending on the shock Mach number ($M_s$), with a critical driver section length, the compressible vortex ring can be a shock-free vortex ring ($M_s < 1.43$), a vortex ring with embedded shock ($1.43 < M_s < 1.61$), or a vortex ring with embedded shock along with counter-rotating vortex rings (CRVRs) ahead of it ($M_s > 1.61$) (Dora et al. 2014, Brouillete & Hebert 1997). Additionally, the deviation of $M_s$ ranges for the classification of a vortex ring with a non-critical driver section is discussed by Dora et al. (2014).

Elder & De Haas (1952) first studied the characteristics of shock-free compressible vortex ring at shock Mach numbers $M_s = 1.12$ and 1.32 experimentally. Baird (1987) identified the presence of embedded shock in the axial region of the vortex ring using differential interferometry. Brouillette & Hebert (1997) first observed the appearance of a tiny vortex ring of opposite circulation to that of the primary vortex ring ahead of it and termed it as CRVR. The primary vortex ring's propagation and dynamics are severely affected by the CRVRs in a highly under-expanded flow regime (Murugan & Das 2010).

Several researchers (Abate & Shyy 2002, Jiang et al. 1997, Sun & Takayama 2003A and 2003B) have studied the diffraction of a shock wave, including the effect of viscous dissipation on it and the time accurate location of the shock is predict6ed. Arakeri et al. (2004) initiated the quantitative experimental study of compressible vortex rings using particle image velocimetry (PIV). Similarly, the study of interactions of a shock wave with a vortex ring (Takayama et al 1993, Tokugawa et al. 1997), and the vortex ring with other generic bodies (Das et al. 2016, Karunakar 2010, Minota, Nishida & Lee 1997, Mishra 2015, Poudel et al. 2018) shows the prominence of compressibility and shocklets in such flows. The formation and evolution of CRVR at high Mach number has been studied in detail along with the effect of driver section lengths and shock Mach number by Murugan & Das (2007A, 2007B, 2008, 2010) using smoke flow visualization. The formation of CRVRs depends on the appearance of the Mach disc in the leading shock cell structure of the trailing jet and its duration of existence for providing necessary impulse to CRVRs (Dora et al. 2014). The study of acoustic characteristics of starting jet and subsequent vortex loop development is important in pulse detonation engines, launch vehicles, and many micro jets studies (Bussing & Pappas 1994). The complete unsteady velocity and vorticity fields along with density, pressure, and temperature fields are essential for understanding the acoustic fluctuations produced during vortex loop formation and evolution. However, such data cannot be extracted from experiments alone. Murugan (2008) and Zare-Behtash et al. (2010) also pointed out that the quantitative data for the compressible vortex core is challenging to obtain from PIV for high magnitudes of circulation.

Non-availability of a measuring system that can simultaneously measure all the physical and thermodynamic properties can be resolved by developing an efficient and accurate numerical tool that can replicate the exact experimental observation as well as is capable of producing simultaneous velocity, pressure, density and temperature field data. Mirels (1955) modelled the laminar boundary layer behind the shock advancing into the stationary fluid. Petersen & Hanson (2003) proposed an improved turbulent boundary-layer model for shock tubes. Ishii et al. (1999) studied the evolution of circular pulsejet using axisymmetric Euler equations with a finite-difference TVD scheme for a wide range of jet strength. They found that the vortex ring near the exit plays a dominant role in the formation of shock cell structure. Murugan et al. (2012) studied the simultaneous velocity, pressure, and density field by numerically simulating a shock tube with driver section length ($D_L$) of 165 mm using the laminar model. They studied the evolution of compressible vortex ring for two distinct flow regimes with PR = 3 and 7. They addressed the translation velocity (U) of the primary vortex ring, which is not measured in most of the studies; although at specific critical instances, U was not estimated accurately. The laminar model was unable to reproduce some complex phenomena, including interactions of vortex rings with each other along with their stretching and deformation. Similarly, the results obtained by Dora et al. (2014) for the compressible vortex ring evolution with PR = 10 and $D_L$ = 285 mm using the numerical simulations with a laminar model was also deficient in imitating the exact time instant for primary vortex ring pinching off and the number of CRVRs formed as per their experimental observation. None of these studies addresses the characteristics of very high Mach number vortex rings.

In this paper, we investigate the development and propagation of compressible vortex rings and their temporal evolutions for varying parameters by implementing a high fidelity turbulence model. The characterization of compressible vortex ring using high fidelity turbulence models has not been attempted before. The present research allows us to shed more light on the underlying physics particularly at high $M_v$, where experiments are not available. The variations of different shock tube parameters result into a change in flow parameters like Reynolds number, peak overpressure and impulse, which shows some interesting phenomena during the formation, evolution and propagation of the compressible vortex ring. Understanding these phenomena are primary focus of this study.

## 2. Numerical details

2.1 Governing equations

From the experiments performed, it is only possible to obtain information about the detailed velocity field and the shock structures. However, to understand the evolution of complicated shock dominated flows, one needs to know the variation of temperature, density and pressure fields (Sun & Takayama 2003B) which are obtained by solving the Favre-filtered (Favre 1983, Soni, Arya & De 2019) Compressible Navier-Stokes equations along with continuity, energy and species transport equations as given below.

Continuity equation:

$$\frac{\partial}{\partial t}(\bar{\rho}) + \frac{\partial}{\partial x_i}(\bar{\rho}\tilde{u}_i) = 0 \qquad (1)$$

Momentum equation:

$$\frac{\partial}{\partial t}(\bar{\rho}\tilde{u}_i) + \frac{\partial}{\partial x_j}(\bar{\rho}\tilde{u}_i\tilde{u}_j) = -\frac{\partial}{\partial x_i}(\bar{P}) + \frac{\partial}{\partial x_j}\left((\mu + \mu_t)\frac{\partial \tilde{u}_i}{\partial x_j}\right) \qquad (2)$$

Energy equation:

$$\frac{\partial}{\partial t}(\bar{\rho}\tilde{E}) + \frac{\partial}{\partial x_i}(\bar{\rho}\tilde{u}_i\tilde{E}) = -\frac{\partial}{\partial x_j}\left(\tilde{u}_j\left(-\tilde{P} + \mu\frac{\partial \tilde{u}_i}{\partial x_j}\right)\right) + \frac{\partial}{\partial x_i}\left(\left(k + \frac{\mu_t C_p}{Pr_t}\right)\frac{\partial \tilde{T}}{\partial x_i}\right) \qquad (3)$$

Species transport equation (evoked for the case with $H_2$ as driver section gas and air in rest of the domain):

$$\frac{\partial}{\partial t}(\bar{\rho}\overline{Y_l}) + \frac{\partial}{\partial x_i}(\bar{\rho}\tilde{u}_i\overline{Y_l}) = -\frac{\partial}{\partial x_j^2}\left(\left(\bar{\rho}D_l + \frac{\overline{\rho u_t}}{Sc_t}\right)\overline{Y_l}\right) \qquad (4)$$

Where, $p = \rho RT$ is the state equation, $\rho$ is the density, $u_i$ is the velocity vector, $p$ is the pressure, $E = e + u_i^2/2$ is the total energy, and $e = h - P/\rho$ is the internal energy and $h$ is enthalpy, $\mu$ is the dynamic viscosity of fluid and $\mu_t$ is turbulent eddy viscosity. Parameters $Pr_t$ and $Sc_t$ are turbulent Prandtl and turbulent Schmidt numbers respectively. In the equations 1-4, the quantities with (-) over them signify that they are Favre averaged and the ones with (~) over them are achieved as a result of the applied filter function. The applied filter in the governing equations ultimately gives rise to sub-grid scale stress (SGS). The subscripts '*i*' and '*j*' are indices for Einstein notations while the subscript '*l*' represents a particular species. In equation 4, $Y_l$ and $D_l$ are mass fraction and mass diffusion coefficient for species '*l*' respectively. Similarly, viscosity ($\mu$) is calculated form Sutherland's law (ANSYS Fluent 2016) with three coefficients as:

$$\mu = \mu_o \left(\frac{T}{T_o}\right)^{\frac{3}{2}} \frac{T_o + S}{T + S} \qquad (5)$$

Where $T$ is the static temperature in Kelvin (K), $T_o$ is reference temperature in K ($T_o$ = 298 K), $\mu_o$ is reference value of viscosity in kg/m-s ($\mu_o$ = 1.849×10$^{-5}$ kg/m-s), S is an effective temperature in K or Sutherland constant.

The present study is based on turbulence model employing detached eddy simulation (DES). In the DES approach, the eddy-viscosity calculation depends on the grid-spacing in the computational domain. Since it is an inbuilt hybrid of LES and RANS approaches, it can be automatically switched to the SGS model. This procedure is carried out by adopting the modification of the dissipation term of the turbulent kinetic energy ($k$) (Soni & De 2018A, 2018B) as follows:

$$Y_k = \rho \beta^* k \omega F_{DES} \tag{6}$$

Where, $\beta^* = 0.09$ is a model constant, $\omega$ is specific dissipation, and $F_{DES}$ is expressed as:

$$F_{DES} = max\left(\frac{L_t}{C_{des}\Delta_{max}}, 1\right) \tag{7}$$

Where, $C_{des}$ is a calibration constant used in the DES model and has a value of 0.61, $\Delta_{max}$ is the maximum local grid spacing $(\Delta_x, \Delta_y)$. $L_t$ is the turbulent length scale that defines the RANS model as follows:

$$L_t = \frac{\sqrt{k}}{\beta^* \omega} \tag{8}$$

One of the problems with the DES formulation is the absence of a mechanism for preventing the limiter from becoming active in the attached portion of the boundary layer. This will happen in regions where the local surface grid spacing is less than the boundary layer thickness, i.e., $\Delta_s < d\delta$, with $d$ of the order one (ANSYS Fluent 2016). To avoid undesirable vortex formation in shear layers, the zonal formulation of DES-based on a blending function of the SST model is implemented. The blending functions are functions of the wall distance. So, the modification in $F_{DES}$ as following is carried out:

$$F_{DES} = max\left(\frac{L_t}{C_{des}\Delta_{max}}(1 - F_{SST}), 1\right) \tag{9}$$

With $F_{SST} = \{0, F_1, F_2\}$ and $F_1, F_2$ are the blending functions of the SST model.

In case $F_{SST}$ is set to 0, the original DES model is recovered. Moreover, $F_{SST} = F_2$ offers the highest level of protection against grid-induced vortex formation, but might be less effective in the LES region (ANSYS Fluent 2016, Kumar, De & Gopalan 2017, Kumar et al. 2016, Soni & De 2018A, 2018B).

2.2 Computational details

The compressible Navier-Stokes equations in cylindrical coordinate along with continuity and energy equations for compressible flows are solved numerically using the CFD solver ANSYS FLUENT (Fluent 2016). A coupled density-based solver is used to solve the equations simultaneously while the DES-based turbulence model is invoked to model viscous effects. Spatial discretization of the convective fluxes is achieved using third-order MUSCL scheme, while the viscous fluxes are calculated using second-order accurate central differencing. The time-stepping scheme is controlled by using the Courant number for stability.

2.3 Computational domain and grid details

Figure 1 depicts the schematic of the computational domain. The domain is such that it represents the actual geometry and scale of the experimental setup used to study the evolution of compressible vortex ring (Dora et al. 2014, Murugan et al. 2012). As the flow field remains axisymmetric in most of its evolution, only one half of the diametral plane is considered in the present simulations with the axis of symmetry along the centreline (line AH of figure 1). The driver and a driven section of the shock tube are represented as Zone 1. The region outside the shock tube is divided in two parts as shown in figure 1 and denoted as Zone 2 and Zone 3. To ensure that the boundary layer is refined along the wall, the grid spacing is distributed hyperbolically along the radial direction represented by Y-axis. The thickness near the wall is decided by keeping y+ value < 5. The mesh is created using ICEM CFD software (Fluent 2016).

2.4 Boundary conditions

The shock tube inner walls (AB, AD, CD) and outer wall (IC) are applied with no-slip conditions. The open boundaries of the domain (IE, EG, GH) are treated by non-reflecting boundary conditions. The non-reflecting boundary conditions are derived based on Euler's equation and can be applied only with pressure outlet conditions. The solution is initialized with ambient pressure 1.01325 bar and temperature 298 K except at the driver section. The pressure at the driver section is applied initially such that the pressure ratios for different cases considered are 3, 7, 8, 10, 12.6, 30, and 50.

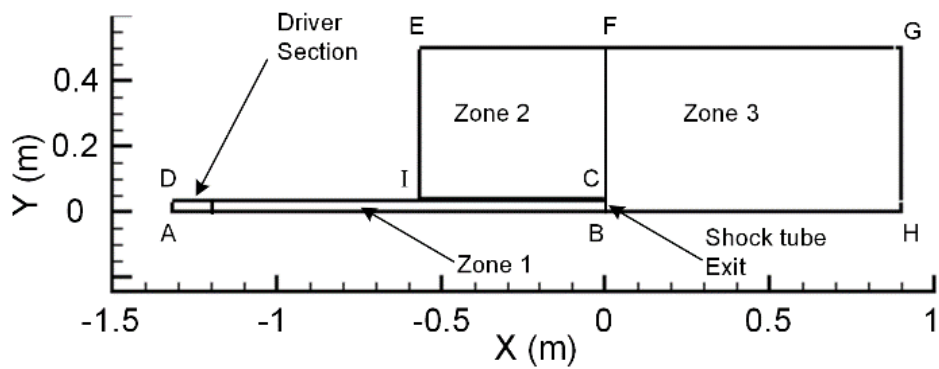

Figure 1. Details of the computational domain

Table 1. Different grids used to obtain a grid-independent solution

|  | Grid 1 | Grid 2 | Grid 3 |
|---|---|---|---|
| No. of Cells | 269,764 | 717,421 | 1,009,938 |

**3. Validation of numerical method**

3.1 Grid independence

Initially, to achieve grid convergence, we perform the simulations using different grids for a particular case with PR = 10 and $D_L$ = 165 mm and air as driver section gas. Table 1 shows the details of 3 different grids considered. By using the results of vortex ring evolution obtained from each case of the grid, Z-vorticity ($\Omega_z$) is extracted along a vertical line $A_1B_1$ (see figure 2a) passing through the center of the vortex core at a same instant t = 850 μs. $\Omega_z$ plot in figure 2b for the case of Grid 1 doesn't show good agreement with the plot obtained from the results of Grids 2 and 3; whereas, there is an excellent match of the same for Grids 2 and 3. Grid 1 could not compute the value of the maximum vorticity at the primary vortex core and other minor variations in its periphery. Similarly, the centreline velocity ($u_c$) of the jet is extracted at the same instant for each grid. Centreline velocity starting from the shock tube exit up to a distance of 0.3 m (i.e. line BH of figure 1) is plotted in figure 2c. It is revealed that the velocity profile for Grid 1 is not consistent with that for Grid 2 at the same instant of time. However, the velocity profiles for the cases of Grid 2 and 3 are in excellent agreement. Jointly, figure 1a-1c demonstrate that Grid 2 is adequate to represent the compressible vortex ring evolution phenomenon.

In addition to this, an analysis employing the Richardson error estimator and Grid Convergence Index (GCI) (Choudhury et al. 2018) is also applied which showed that the error value for the fine grid is reasonably lower than that for the coarse grid. The detailed analysis is given in Appendix A1. Finally, the optimum mesh size of ~0.7 million (Grid 2) is considered for all further simulations hereafter.

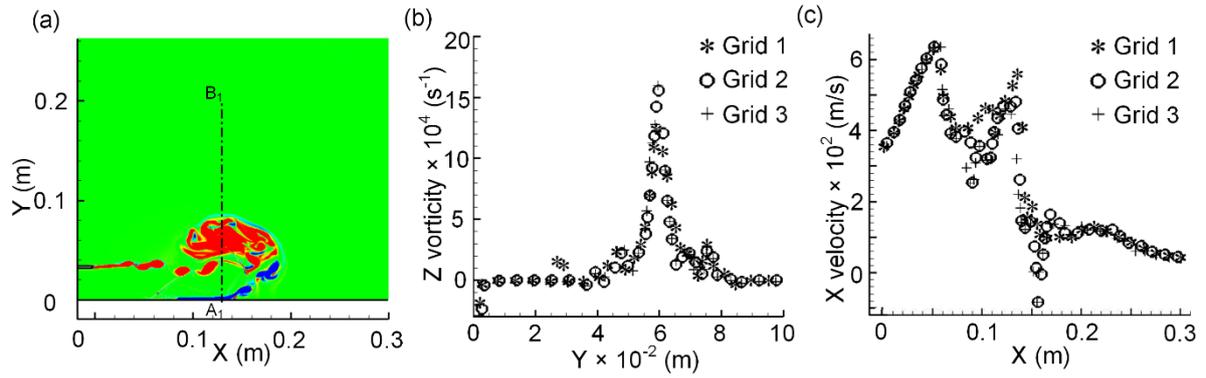

Figure 2. Illustration of grid-independence test considering vortex ring evolution with PR = 10, $D_L$ = 165 mm (a) Vortex ring core with line $A_1B_1$ passing through the center, (b) Z-Vorticity distribution through the line $A_1B_1$, (c) Centreline velocity profile of the jet

3.2 Effect of viscous model

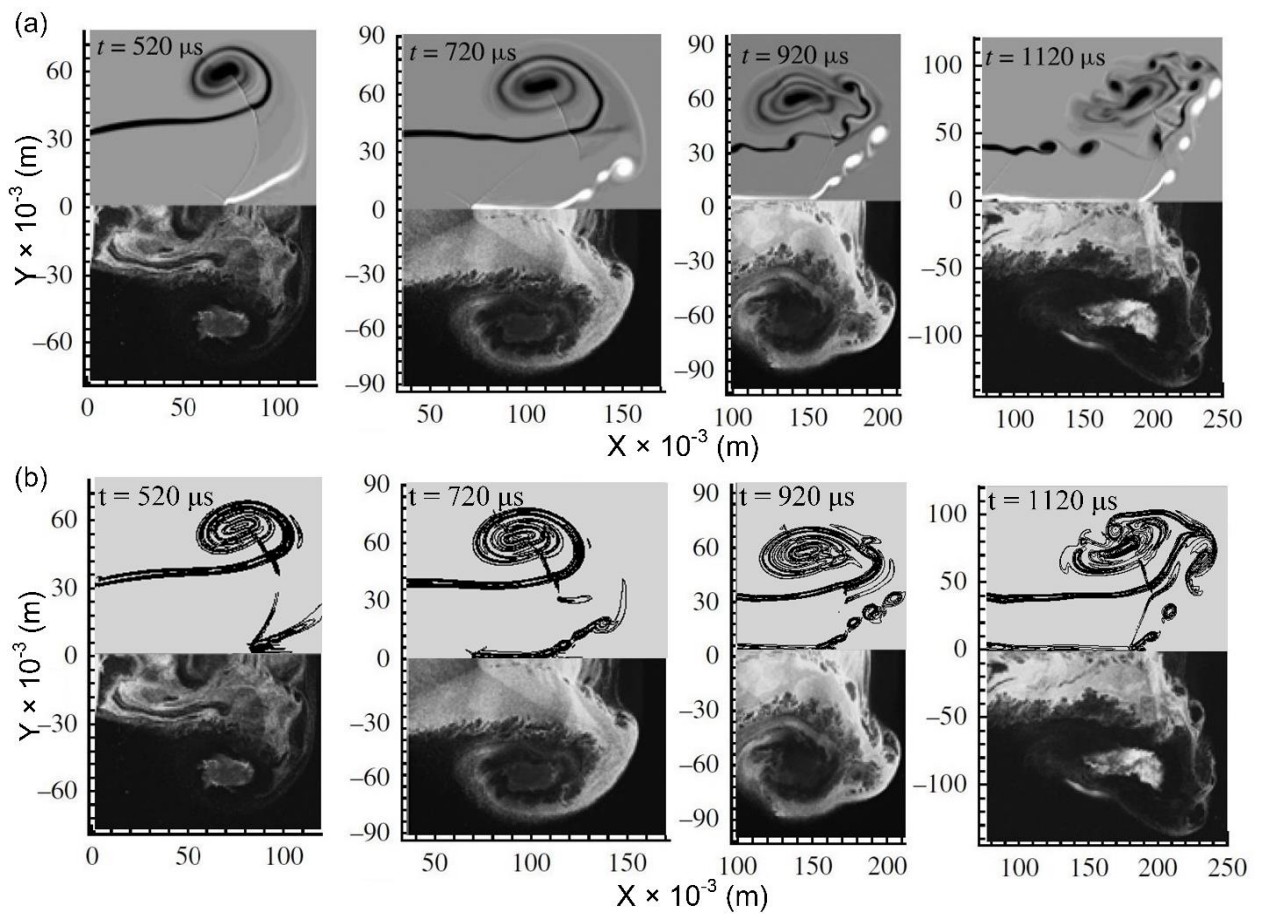

Figure 3. The temporal and spatial resolution of vortices achieved in the present study with DES simulation as compared to the previous study with laminar simulation (Dora et al. 2014). (a) Experimental observation using smoke flow visualization (Dora et al. 2014) (bottom half) versus the vorticity plot obtained through laminar

simulation (top half) (Dora et al. 2014) and (b) Experimental observation using smoke flow visualization (Dora et al. 2014) (bottom half) versus the vorticity plot obtained in the present study (top half).

Accurate determination of characteristics of a vortex ring at very high PR is difficult experimentally whereas, it is feasible numerically. Such a vortex ring is highly turbulent and incorporation of an appropriate turbulent model is required. To establish the effect of the turbulence model first, the simulation has been carried out and compared with the previously published experimental results (Dora et al. 2014) for PR = 10 and $D_L$ = 285 mm. Figure 3 depicts the vorticity contours obtained from laminar simulation (Dora et al. 2014) and DES simulation respectively at different times in association with the flow visualization images (Dora et al. 2014). It is observed that the DES model reproduces the flow features more accurately as compared to the laminar model (Dora et al. 2014). At t = 920 μs the number of CRVRs predicted by the laminar model (Dora et al. 2014) is three (see figure 3a), however, the present simulation shows four CRVRs (see figure 3b) which is consistent with the experimental observations of Dora et al. (2014). Additionally, the laminar model suggests that multiple shear layer vortices start to form at t = 920 μs, in contrary to t = 720 μs seen in the experiment and the present simulation. The present simulation also captures the merging of two slipstream vortices at this instant. Also, the smeared structure of the shear layer in experiments (Dora et al. 2014) is an indication of the turbulent nature of the shear layer whereas, the laminar solution shows discrete vortices (Dora et al. 2014) at t = 1120 μs. Thus, the present simulation with the DES model is expected to provide accurate results at even higher PR cases where the effects of CRVRs and slipstream vortices are vital.

3.3 Ring characteristics validation

In order to gain further confidence in the chosen SST-k-ω based DES model, in this section, we have compared the results of temporal variation of vortex ring characteristics for the case of PR = 7 with the published results of Murugan et al. (Murugan et al. 2012) as demonstrated in figure 4. A sequence of vorticity snapshots shown in figure 4a illustrates the evolution of the primary vortex ring along with its interaction with the shear layer vortices. Figure 4b depicts the variation of non-dimensional translation velocity with time for this particular case obtained in the present study as compared to the experimental as well as laminar simulation results of Murugan et al. (Murugan et al. 2012). The assertion of our current viscous model (SST-k-ω based DES) being better than the laminar model is well justified as it predicts the non-dimensional velocity more accurately than the previously published numerical study (Murugan et al. 2012). Especially at specific critical points, the laminar model could not reproduce the results of experiments because some complex phenomena including the interaction of primary vortex ring with shear layer vortices and other small eddies take place along with stretching and deformation of them. These events severely affect the translation of the primary vortex ring and require effective turbulent modelling to capture the data accurately.

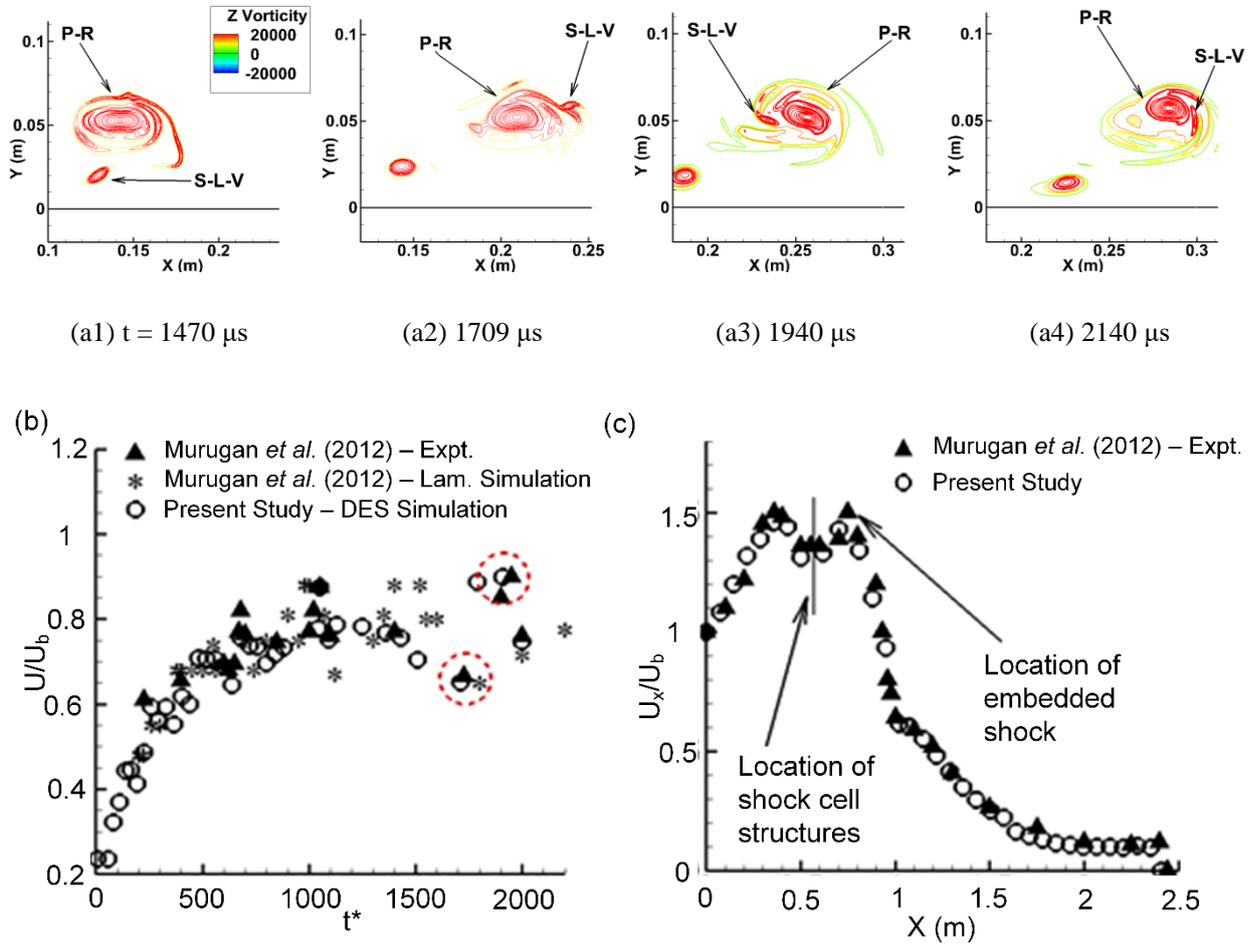

(a1) t = 1470 μs    (a2) 1709 μs    (a3) 1940 μs    (a4) 2140 μs

Figure 4. Validation of present DES based simulation with experiments (Murugan et al. 2012) and laminar simulations (Murugan et al. 2012) for vortex ring evolution at PR = 7. (a) Vorticity contour for PR = 7 showing the interaction of primary vortex ring with shear layer vortex at different instants, (b) Variation of non-dimensional translation velocity of vortex ring, (c) Non-dimensional centre line X-velocity ($U_x$) of jet at t = 560 μs

Moreover, the sudden jump in translation velocity of the ring from t = 1709 μs to t = 1940 μs (indicated by the red dotted circles in figure 4b which is observed both in the present study and the experiment of Murugan et al. (2012) is due to the effect of a shear layer vortices on the primary vortex ring. At t = 1709 μs (see figure 4a-2) the shear layer vortex intensifies the axial velocity of the primary vortex ring as it is in front of it; As the former encircles and travels to the upstream direction of the latter at t = 1940 μs (see figure 4a-3) the translation velocity declines. Additionally, the centreline velocity profile of the jet obtained from the present study is also in a good agreement with the experiment (Murugan et al. 2012). It is inferred that at later instants of time, shear layer vortices (see figure 4a-4) merges with the primary vortex ring core. This further demonstrates the necessity of the turbulence model in the accuracy of the simulations and thus, resolving the small scale vortices. The rest of the results reported subsequently are using SST-k-ω based DES model only, unless specified.

**4. Results and Discussion**

The dynamics of compressible vortex rings in different flow regimes are illustrated through the characteristics of velocity and vorticity field with quantitative prediction of their geometric and kinematic characteristics. Based on flow features, the regimes can be delineated as (i) shock free vortex ring (here, PR = 3 case), (ii) vortex ring with embedded shock (PR = 8 case), (iii) vortex ring with embedded shock, Mach disc and CRVRs (cases considered for PR > 10) and (iv) very high shock Mach number vortex rings (cases considered with air and hydrogen as driver section gas for PR = 50). Although, compressible vortex rings in first three regimes had been studied earlier, the flow dynamics in regime-iv illustrate many new features of such flows that were hitherto unknown. It is to be noted that at such a high PR, sufficiently high impulse is produced to generate a vortex ring translating at supersonic speed. Examples in all range of Mach numbers have been discussed to illustrate the comprehensive picture of such flows. Parameters considered for different cases are shown in table 2. The flow dynamics observed during the vortex ring evolution with PRs 12.6 and 30 is provided in section A2 of the Appendix.

Variation of vortex ring parameters such as vortex ring diameter (D), translation velocity (U), circulation (Γ) and vortex core diameter ($D_c$) with time (t) have been studied for all cases considered. The parameters are non-dimensionalized as D/d, U/$U_b$, Γ/($U_b$·d), $D_c$/d and t* = (t·$U_b$)/d respectively. Here, $U_b$, $ρ_b$, and $ν_b$ are respectively the velocity, density and kinematic-viscosity behind the non-diffracted shock wave, i.e. while the shock is at the exit of the shock tube and $P_r$ is the pressure ratio across the shock at the same instant. The diameter of the shock tube is denoted by d (= 0.064 m). The time when the incident shock is at the shock tube exit is taken as t* = 0 (t = 0 μs). X-Y coordinate system has been used to depict the plots. Here, X-axis represents the axial direction of the vortex ring while Y-axis represents the radial direction, which is consistent with the computational domain depicted in figure 1.

Table 2. Variation of flow parameters for different cases considered

| PR | $M_s$ | $M_v$ | $d_o$ (mm) | $D_L$ (mm) | Gas | $U_b$ (m/s) | $P_b$ (Pa) | $T_b$ (K) | $\rho_b$ (kg/m$^3$) | $v_b$ (kg/m.s) | $P_r$ |
|---|---|---|---|---|---|---|---|---|---|---|---|
| 3 | 1.3 | 0.31 | 64 | 165 | Air | 132.57 | 168428 | 348.124 | 1.68555 | 1.9 E-5 | 1.66 |
| 8 | 1.55 | 0.51 | 64 | 165 | Air | 235 | 244628 | 392.46 | 2.1647 | 2.1 E-5 | 2.414 |
| 10 | 1.61 | 0.53 | 64 | 119 | Air | 279.14 | 282508 | 414.457 | 2.375 | 2.35 E-5 | 2.78 |
| 10 | 1.61 | 0.53 | 64 | 165 | Air | 279 | 283178 | 414.51 | 2.38 | 2.35 E-5 | 2.79 |
| 10 | 1.61 | 0.53 | 64 | 285 | Air | 279.02 | 282449 | 414.43 | 2.374 | 2.35 E-5 | 2.79 |
| 12.6 | 1.68 | 0.54 | 64 | 165 | Air | 308.03 | 310648 | 429.406 | 2.5204 | 2.43 E-5 | 3.06 |
| 30 | 2.75 | 0.60 | 64 | 165 | Air | 415.35 | 430779 | 492.584 | 3.047 | 2.65 E-5 | 4.25 |
| 50 | 2.92 | 0.63 | 64 | 165 | Air | 478.512 | 512110 | 535.42 | 3.33219 | 2.75 E-5 | 5.05 |
| 50 | 2.28 | 1.08 | 64 | 165 | H$_2$ | 1042 | 1.520E6 | 211.722 | 1.7297 | 0.84-5 | 15 |

4.1 Shock tube exit flow conditions

The shock tube exit flow parameters for all the cases considered are shown in figure 5. The parameters: exit pressure and velocity history with time that are shown in figures 5a and 5b respectively drives the flow. From table 2 we observe three-fold increase of the exit pressure ratio ($P_r$) at the time of arrival of the shock at exit when PR=50 cases for H$_2$ and Air are compared. Thus, comprehensive understanding of the exit flow parameters is more important than the classification based on PR and $D_L$. Figure 5a shows the effect of PRs as well as $D_L$ on the exit flow pressure. This exit flow pressure, in turn, shows the nature of expansion occurring at the shock tube exit due to shock diffraction. Similarly, figure 5b characterizes the temporal variation of the axial component of the exit flow velocity for different cases considered. The exit velocity increases due to entry of expansion waves into the tube when $U_b$ is subsonic. However, for PR=30 and 50 cases the exit velocities remain constant for certain time as the flow is supersonic and does not allow the propagation of expansion waves inside the shock tube. Additionally, it is demonstrated later that, exit conditions are useful in computing the jet impulse as well as peak-over-pressure, which in turn sheds light on the criteria of formation of vortices of opposite circulation or counter-rotating vortex ring (CRVR) at high PRs.

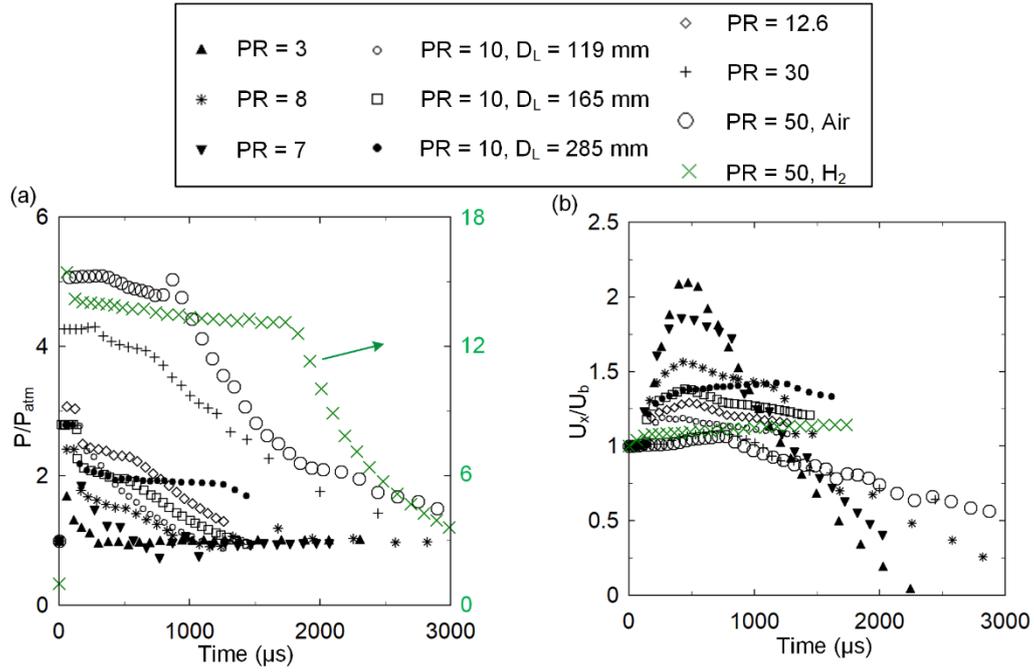

Figure 5. Time variation of non-dimensionalized (a) exit pressure and (b) exit flow axial velocity

4.2 Evolution of vortex ring for PR = 3 (subsonic shock-free vortex ring)

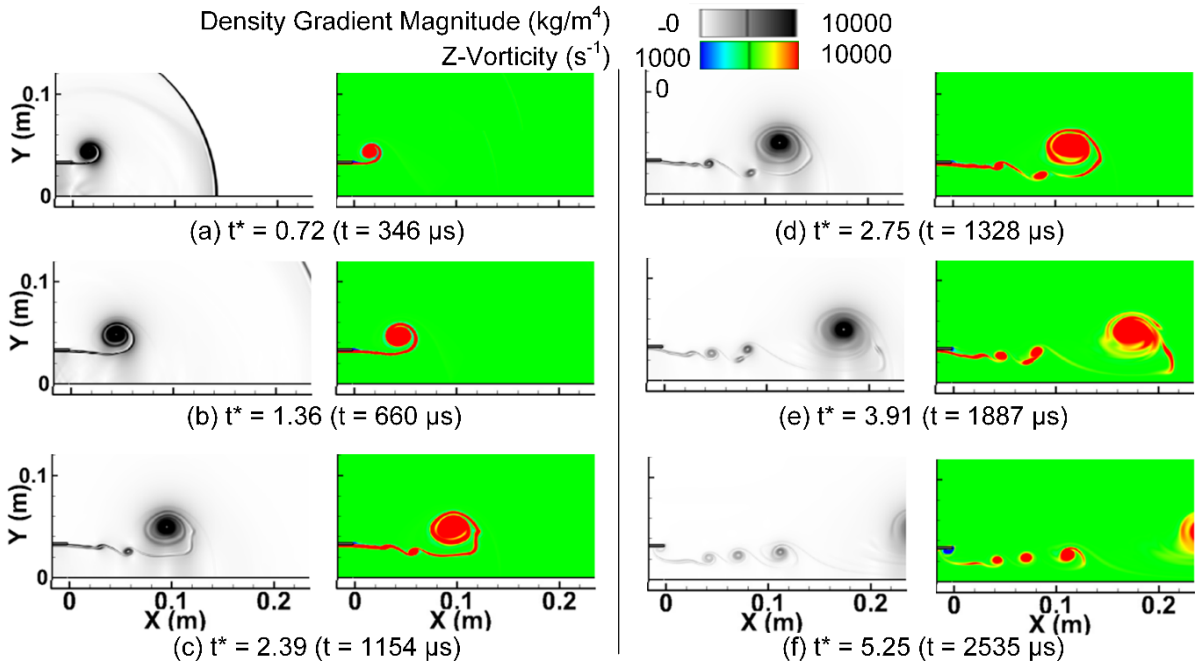

Figure 6. Density gradient magnitude plot and vorticity contour plot of the evolution of compressible vortex ring at PR = 3.

Figure 6 demonstrates the evolution of the vortex ring for PR = 3, $M_s$ =1.3. The vortex ring grows during the initial roll-up process and moves in both the radial as well as the axial direction. As the pressure inside the shock tube is higher than the ambient pressure and flow behind the shock is subsonic ($U_b$ = 132 m/s), disturbances propagate into the shock tube and accelerates the flow inside it (Murugan et al. 2012). It has also been observed experimentally and numerically that the jet flow remains subsonic (Ishii et al. 1999) for PR < 4.1 in case of air and the vorticity fields and numerical shadowgraphs (density gradient magnitude plots) do not show any shock cell structures or any significant density variation in the central region of the vortex ring (see figure 6).

The shear layer vortices enter the vortex core as the jet continues to feed vorticity into the vortex ring and thus reinforces its strength (Figure 6). As the strength of the vortex ring grows, the self-induced velocity of the ring also increases. At t* = 1.36 (t = 660μs), the jet boundary shrinks inwards as the trailing jet velocity becomes slower than the vortex ring's translational velocity. Figure 6c exhibits the vortex ring at t* = 2.39 (t = 1154μs), where the formation of shear layer vortices initiates and the vortex ring continues to propagate towards downstream. The shear layer vortices formed at the trailing jet is due to Kelvin-Helmholtz type instability and grow faster in subsequent instants of time. The trailing jet feeds vorticity into the primary vortex ring via shear layer vortex as observed at time instant t* = 3.91 (t = 1887μs). Subsequently, there is no more feeding of vorticity into the primary vortex ring; thus, the primary vortex ring is said to be pinched-off. The shear layer vortices become distinct isolated structures of gradually diminishing size as seen from figure 6f. The effect of trailing jet on the fully formed vortex ring is not significant as the flow expands like a subsonic jet (Lumley 1967).

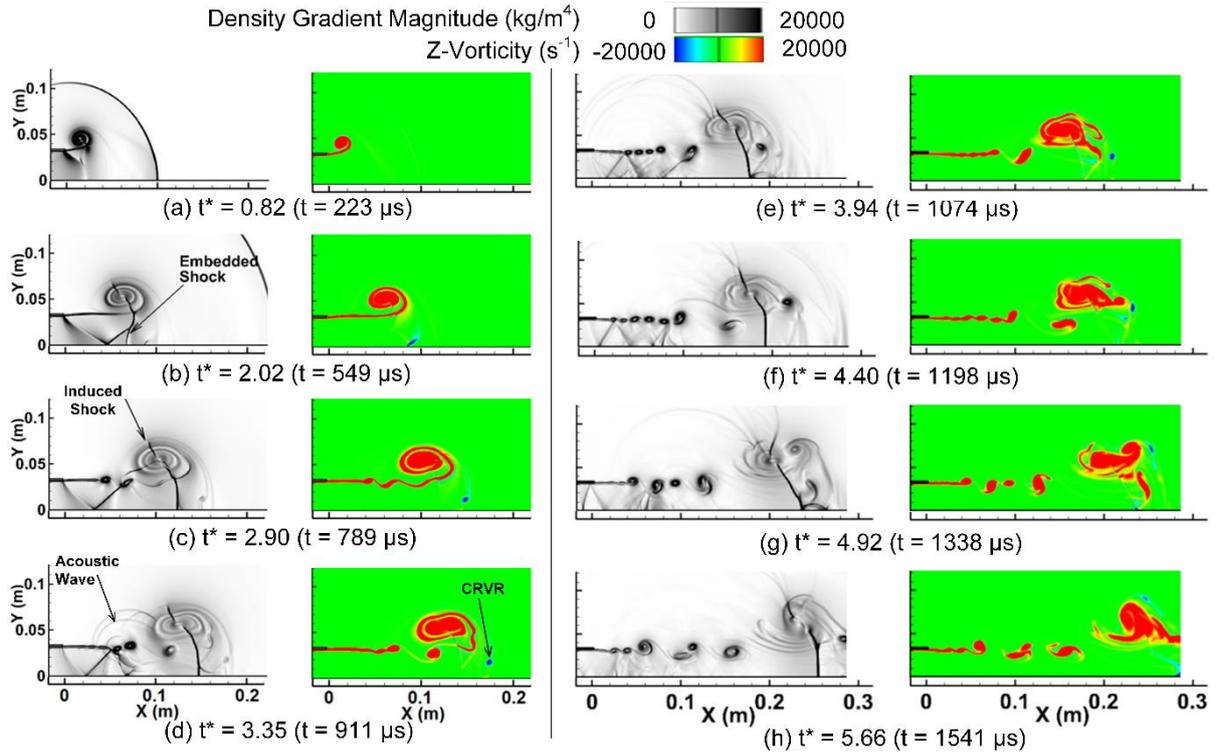

Figure 7. Density gradient magnitude plot and vorticity contour plot of the evolution of compressible vortex ring at PR = 8.

4.3. Evolution of vortex ring for PR = 8 (vortex ring with embedded shock and CRVR)

The complete formation and growth of the primary vortex ring for PR = 8 are presented in figure 7. At this PR, the disturbance waves continuously propagate inside the shock tube until the flow reaches sonic conditions. Since the pressure ratio across the incident shock at the exit of the shock tube is more than *1.89*, the flow develops like an under-expanded jet. The value, *1.89* is calculated using the inviscid choked flow condition for air. An embedded shock near the central region and a vortex induced shock in the outer periphery of the vortex core is distinctly observed similar to earlier studies (Dora et al. 2014, Murugan et al. 2012) in the numerical shadowgraph of figure 7b. It is also found that the vortex core is stretched and not circular anymore. It is interesting to note that as the PR increases the vortex core shape becomes more elliptical (also see figure A1 in Appendix A2 for PR =12.6 and 30 cases)

Numerical shadowgraph also confirms the shock cell structure formed in the axial region. At t* = 2.90 (figure 7c, t = 789 µs), formation of shear layer vortices initiates. From the density gradients of figure 7 c, d it is clear that discrete shear layer vortices are formed when the oblique shock of the shock-cell structure interacts with the shear layer. Emission of shock-vortex noise is also clear from figure 7d. Merging of the shear layer vortices and formation of stronger spiralling vortex-pair structures are observed in the entire process (see figure 7e - 7h) similar to incompressible vortex ring (Widnall & Sullivan 1973, Bhatia & De 2019). The Biot-Savart interaction of primary and shear layer vortex rings results in increase in diameter and decrease in velocity of the primary (can also be seen in figure 9a and 12b for $t^* > 4$). Two important dynamics are noted during the interaction which is different from lower PR cases. As the merged vortex crosses the embedded shock it gets stretched. The stretched merged vortex is sufficiently strong to leapfrog slightly ahead of the primary before being engulfed into the core of the primary. It is to be noted that, at this PR, the interaction of these shear layer vortices with the primary vortex ring and consequently engulfing of former into the latter is observed as long as the primary vortex ring crosses the domain of computation, i.e. until the time of observation. These observations indicated that even though it appears that the vortex ring has pinched-off from the trailing jet at $t^* = 3.94$ (figure 7e) however, primary vortex is continued to feed by the discrete shear layer vortices. In this case, only a tiny CRVR forms as the Mach disc appears for a short duration (Dora et al. 2014) during which the exit pressure ratio is greater than 1.8 (see figure 5).

Multiple CRVRs form as the PR is increased and the duration of exit Pr stays as more than 1.8 for a longer duration as observed in section A2 of the Appendix. The discussion is not being included for brevity and also already available in the article by Dora et al. (2014). The only notable difference is in the interaction of the shear layer vortices with primary vortex ring in the presence of CRVRs. As the CRVRs become stronger (say PR =12.6, figure A1), it accelerates the shear layer vortices due to induced velocity and squeezes them between primary vortex ring and the CRVRs and thus, accelerate their entrainment into the primary core. However, for very high PR, the characteristics of such compressible vortex rings are quite different and experimental study is not available and is considered next.

4.4 Evolution of vortex ring for PR = 50 (air as driver section gas)

In PR = 50 (air) case Figure 8, due to the diffraction of the high $M_s$ shock and rapid expansion of flow at the exit, the vortex ring and trailing jet shear layer starts moving rapidly in the radial direction. The radial motion of the vortex ring for this case is significantly higher than that is observed in low Mach number cases. In the initial phase $t^* < 3$, the barrel shock formation and its interaction with the embedded shock make the Mach disc diameter quite large (figure 8 a, b). It pushes the shear layer and vortex ring radially outwards which results in a curved shear layer that has not been observed at low PR where, the barrel shock is not formed.

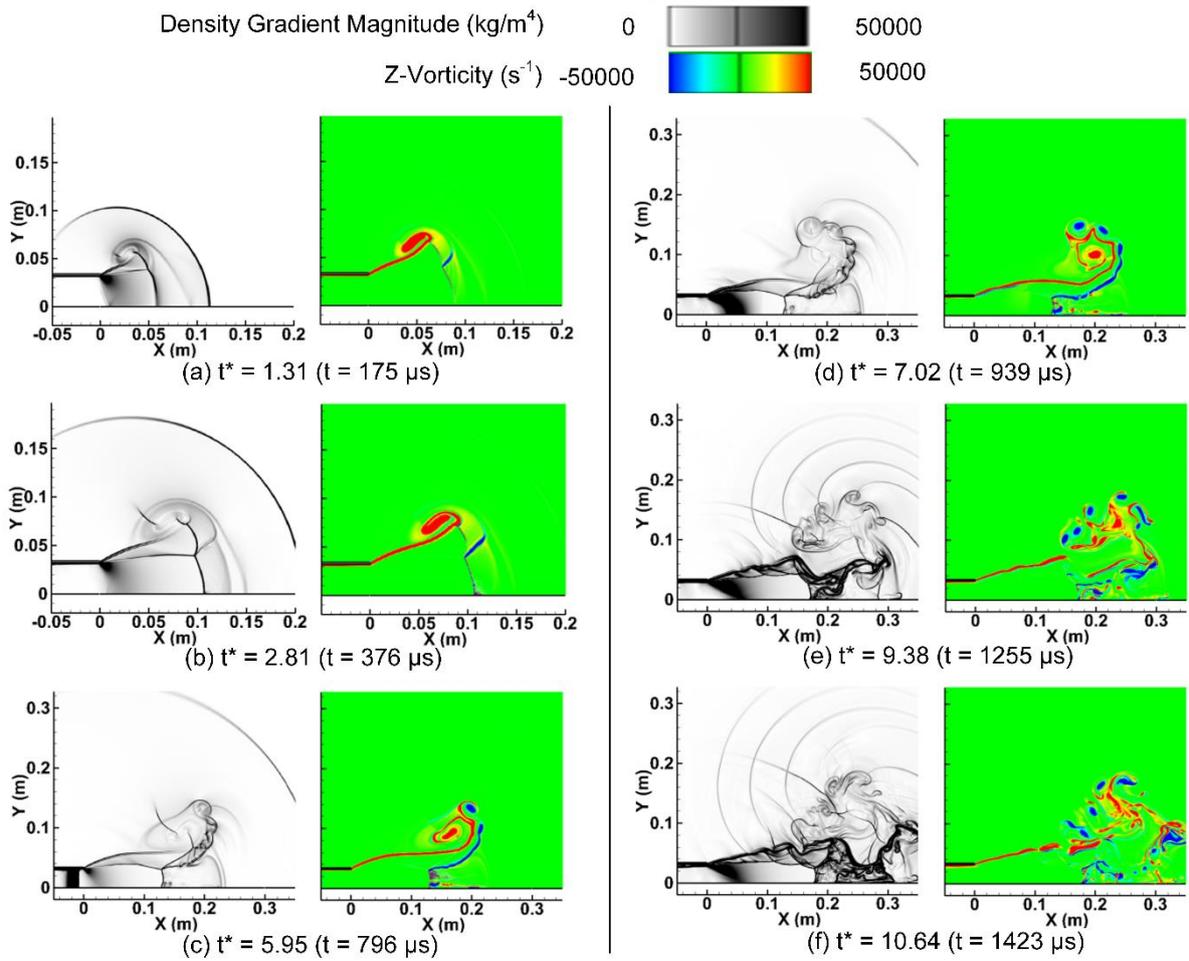

Figure 8. Density gradient magnitude plot and vorticity contour plot of the evolution of compressible vortex ring at PR = 50.

To illustrate and understand why it happens and further differences between low and high PR cases, the flow dynamics of two intermediate cases (PR =12.6 and PR =30) are considered (A2 of the Appendix). It can be observed that the shock-cell structure of PR =12.6 is different from that of PR = 30 where the barrel shock is formed. At the lower PR case the Mack disc before disappearance is very small and the triple point stays very close to the axis (see figure A1c - A1d) whereas, the triple point forms away from the axis at high PR cases (see figure A1k - A1l and figure 8f). The location of the triple point is strongly dependent on PR or $M_s$. In case of PR = 12.6 and 30, the triple point initially stands at a radial distance of 0.018m and 0.034m at $t^* = 1.58$ (t = 330μs) and $t^* = 2.20$ (t = 338μs) respectively (see section A2 of Appendix) as compared to 0.040m at $t^* = 2.81$ (t = 376μs) in case of PR = 50 (figure 8b). To satisfy the conditions imposed by the primary vortex ahead of the triple point, another shock is observed to form at the triple point (figure A1i - A1j), which eventually creates multiple shocks and changes the structure of the curved barrel shock (figure 8c).

The slipstream from the triple point grows and rolls into CRVRs as observed by Dora et al. (2014). The interaction of the primary vortex ring and CRVR is responsible for the further radial growth of the vortex ring (figure 8e). CRVRs because of its opposite circulation have self-induced velocity towards the shock tube exit and rolls around the periphery of the primary vortex ring. The subsequent interaction of multiple CRVRs, slip-stream shear layer with primary vortex ring results in the breakdown of the core at around t* =10 (figure 8 e, f). It can also be observed that the Mack disc disappears around this time (figure 8e) and the generation of slipstream vortices consequently ceases.

Another important difference between low and high PR cases is in the number of CRVRs that are formed. In the case of PR = 8, only one weak, tiny CRVR forms (figure 7b - 7d) whereas, multiple CRVRs are formed for high PR cases (figure 8 and section A2 of the Appendix). The number of CRVRs formed for different parameters of PRs and $D_L$ in association with the jet impulse is discussed in detail in section 4.7.

It is noteworthy to mention that the CRVRs interact strongly with the primary vortex ring. Figure 8d shows that the primary vortex ring is severely affected by the peripheral motion of multiple CRVRs. It is also observed that the primary vortex core becomes turbulent and unstable, which almost losses its identity. Murugan and Das (2008) also identified this phenomenon of vortex ring transiting to a turbulent structure before it pinches-off from the trailing jet for Ms > 1.5. The turbulent nature of the vortex ring due to interaction with strong CRVRs is also evident in cases of PR = 12.6 and 30 (figure A1 of the Appendix).

An interesting feature observed in the case of PR = 50 is the formation of a second triple point close to the axis in addition to the usual slipstream formed at the triple point away from it. The slipstream from the second triple point as seen in figure 8d latter grows into multiple tiny vortices of the same nature as CRVRs. These vortices latter merge with stronger CRVRs, as observed in figure 8 e, f. Here, some tiny vortices of same circulation of primary ring is also observed. It is not very clear why these vortices are formed. However, when the exit impulse is increased further with $H_2$ as the driver section gas as PR=50, these vortices are more prominent. Characteristics of PR 50 $H_2$ as driver gas cases are discussed next.

4.5 Compressible Vortex Ring at PR = 50 and Hydrogen as Driver Section Gas

It is difficult to obtain a vortex ring translating at supersonic speed using air as driver section gas as observed in Table 2. With air as driver gas, when the pressure ratio is raised from 30 to 50, the $M_v$ only increases marginally from 0.6 to 0.63. Thus, a substantial increase in exit impulse is required to produce a vortex ring that may travel at a speed of greater than sound speed. Hence, to achieve $M_v$ greater than one, $H_2$ is used as driver gas with PR = 50. Other conditions are kept the same as PR = 50 air case

The initial formation process for PR =50, $H_2$ case (figure 9) is quite different than the earlier cases discussed. The rapid entrainment caused by the roll-up of the separated shear layer from inside the shock tube creates a flow over the outer peripheral surface of the shock tube near the exit. This opposite circulation vorticity layer wraps up the vorticity layer that rolls up into the primary vortex ring as seen in early times t* = 0.49 (t = 30 μs). The interesting feature here, is the instability of this reverse vortex layer (RVL) in the corner region and creating additional tiny vortices in subsequent times t* = 1.46 – 10.54 (t = 90 μs – 647 μs). The numerical shadowgraph for the corresponding case is shown in figure 9 and shows the nature of the shock structure as well as shocklets in the flow field. It also illustrates that these vortices are considerable sources of noise.

Another interesting observation is the distortion of primary vortex during formation at t* = 0.49 (t = 30 μs). A portion of the primary vortex is stretched out during the formation of the vortex shock and the barrel shock in this unsteady under expanded flow. The expansion of the flow behind the expanding incident shock stretches this vortex layer towards the centre (figure 9 b, c). The slipstream originated from the triple point separates this stretched vertical region into two halves (figure 9c), one part of which gets merged with the primary vortex core and the other half develops into an interesting structure. The vortex layer stretched out in the region ahead of the Mach disc (see figure 9a) is not observed in the earlier cases.

The formation of the much stronger slipstream and CRVRs again encircle the primary ring and distorts it completely. The CRVRs and RVL vortices are indistinguishable at this stage (see figure 9g). The stretched layer of vortical fluid at this stage is pulled towards the centreline and forms a wavy structure (see figure 9 f, g). It is also observed that the curve Mach disc changes its shape to satisfy the symmetry condition at the centre and conditions ahead of it and forms another triple point near the centre, which, creates a tiny vortex of circulation of CRVRs which stays and grows as its self-induced velocity is upstream. At this point near the centre a type IV shock-shock interaction type structure forms and an opposite circulation (i.e., the circulation of primary vortex ring) vortex layer forms (see figure 9f). This vortex layer quickly breaks down into tiny vortices (TV) and starts moving rapidly forward. While the TVs move rapidly downstream and cross the primary vortex zone near the centre, the opposite circulation tiny vortex remains in this region and grows slowly. This opposite circulation vortex due to forming a Mach reflection like structure near the centre is also observed in the case of PR 50 with air as driver section gas. For comparison, these vortices are shown in figure 10 where the numerical values of the density gradient magnitude is normalized with the corresponding maximum value. To illustrate the unique kind of shock waves, density field, and flow dynamics demonstrating for $H_2$ case, figure 10 is constructed by taking the mirror image of the axisymmetric half portion. The density field also shows the rapidly moving TVs near the centre. Another interesting observation (see figure 17e) in the density field is, a clear density structure just below the primary slipstream that forms the CRVRs. This vortex is not so visible in the vorticity field. However, careful observations show that the clear density structure is formed by the stretched out vortex layer as it intensifies.

Having noted down the detailed flow, shock and vortex structures and their variation for low to high PR cases we further looked into the implications of these phenomena on the geometric and kinematic characteristics of the vortex rings. In particular, how vortex ring diameter, core diameters, circulation and translational velocities are different for high $M_v$ vortex rings compared to low $M_v$ cases are quantitatively investigated.

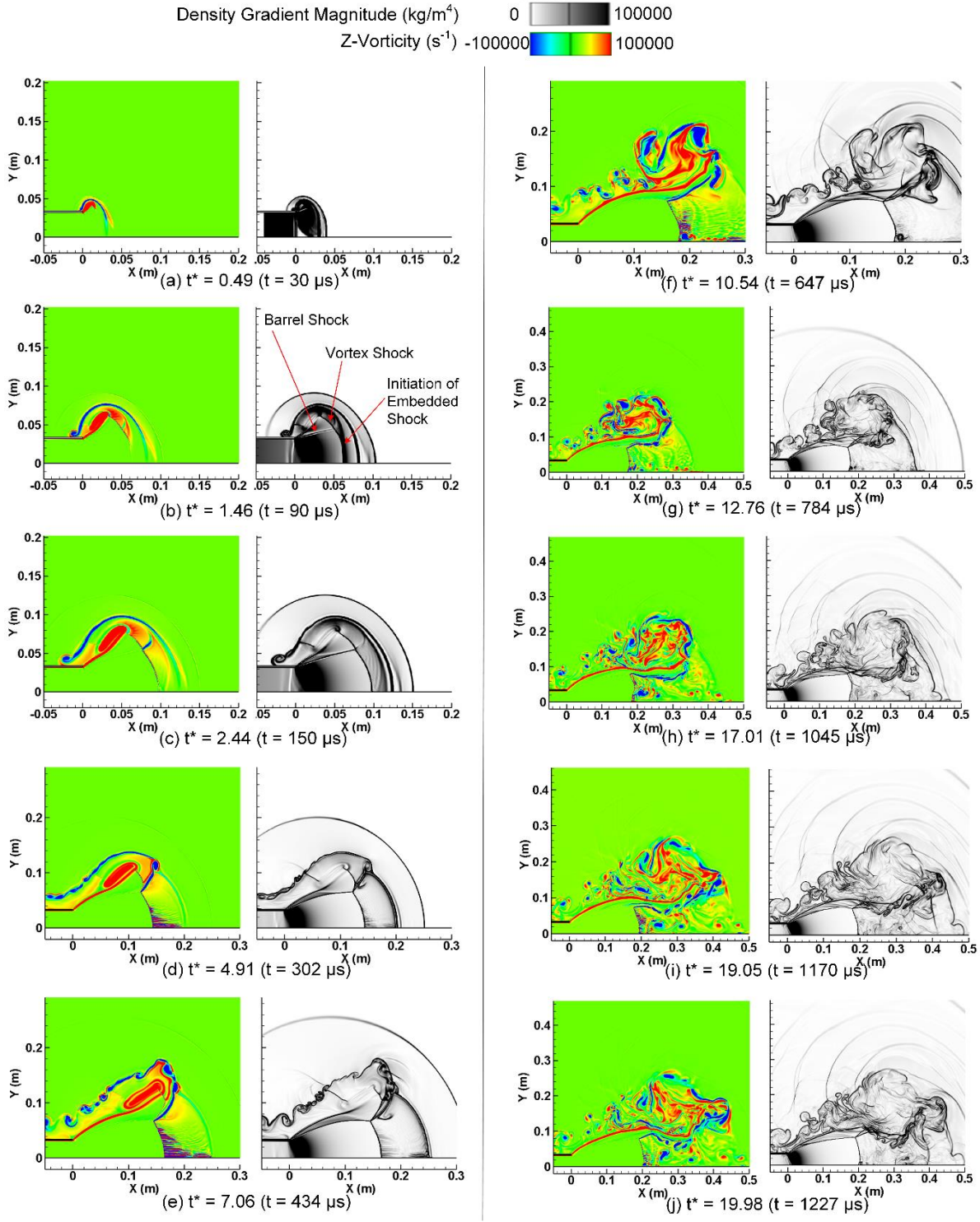

Figure 9. Evolution of compressible vortex ring with time illustrated through snapshots of numerical shadowgraphs and vorticity contour plot for the case with PR = 50 and $H_2$ as driver section gas.

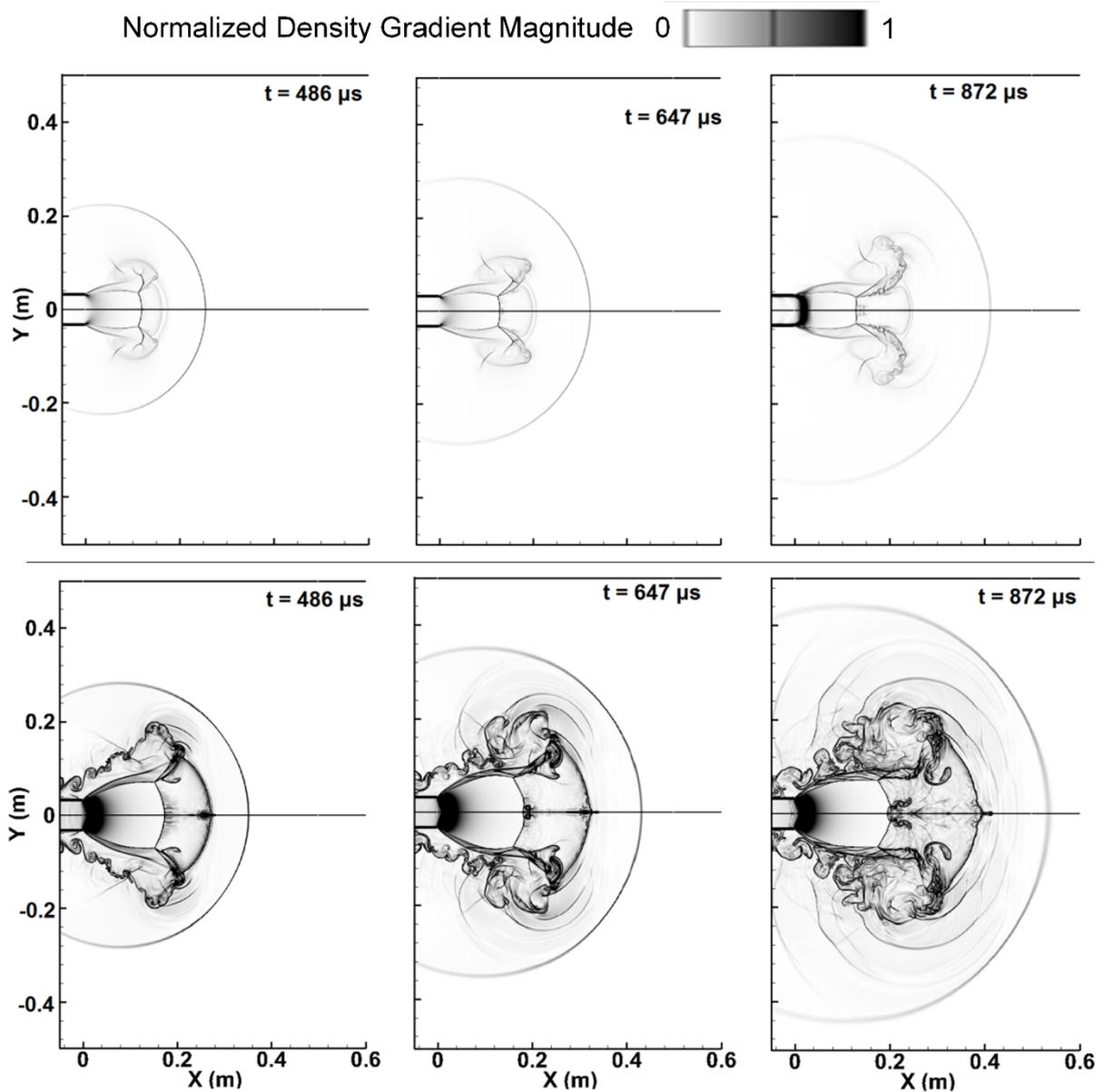

Figure 10. Time-to-time comparison of the shock waves and flow features for vortex ring propagation with driver section gases air (top row) and hydrogen (bottom row) at PR = 50.

4.6 Characteristics of Vortex Ring

Ring diameter and core diameter are two essential geometric parameters that describes a vortex ring along with circulation and translation velocity, the other two kinematic parameters that characterizes its motion. Variations of all these parameters with time for different Mach numbers based flow regimes illustrate the implications of complex evolution process on compressible vortex rings.

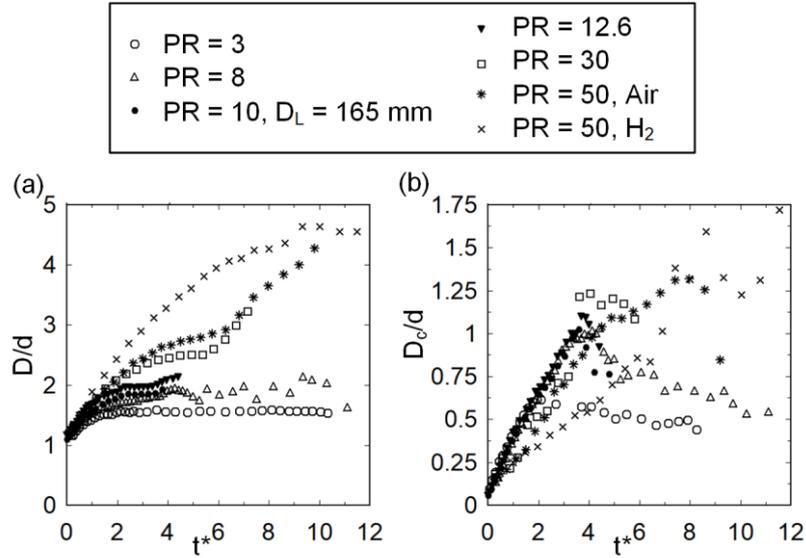

Figure 11. Evolution of compressible vortex ring characteristics with time (a) Non-dimensional vortex ring diameter (b) Non-dimensional vortex core diameter.

4.6.1 Ring Diameter (D)

Figure 11a illustrates the variation of vortex ring diameter with time for all PR cases corresponding to three different flow regimes. The embedded shock-free vortex ring (for PR = 3) diameter is found to increase initially during the formation time and then attain constant value at approximately $t^* = 1.24$ (t = 600 μs), similar to an incompressible vortex rings free of stopping vortex (Das et al. 2017). After reaching a constant diameter in PR = 8 case, a marginal increase in ring diameter is observed at $t^* = 4.40$ (t = 1198 μs) due to the induced effect of CRVR as it moves ahead of the primary core (Murugan & Das 2010). The Primary ring diameter reduces marginally when the CRVR moves behind the core-centre of the primary ring due to the same induced effect, which also accelerates the flow (Murugan & Das 2010). The leapfrogging (Maxworthy 1972, Riley Stevens 1993) action of the shear layer vortices has the opposite effect and is evident from the vorticity field and numerical shadowgraphs of PR = 50 (see figure 8f). For higher PR cases (PR = 30 - 50), the ring diameter continues to increase due to interaction with slipstream and as multiple CRVRs. As an example for the PR = 50, air case, the slipstream CRVRs ascends gradually up to $t^* = 6.36$ (t = 850 μs), followed by a rapid growth when multiple CRVRs are formed. It is to be noted that the slipstream itself also pushes the primary ring outwards. For PR 50 $H_2$ case, the trailing jet shear layer vortices, primary vortex ring and CRVRs have relatively higher strength and hence, the ring diameter for hydrogen is greater than other cases considered. As observed in figure 11a, in case of air, the vortex ring diameter reaches to an almost steady value (~170 mm) and at $t^* = 6.28$ (t = 840 μs), after which, multiple CRVRs causes the rapid radial growth of the diameter. For hydrogen case, the corresponding effect is observed much later at t* ~ 10, and the ring diameter shoot up after t* ~ 10 also perceived through figure 9 f, g.

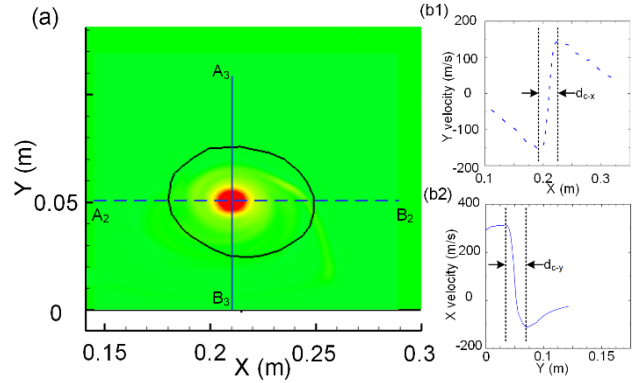

Figure 12. A portion of the azimuthal plane (XY plane) depicting vortex core with (a) a closed-loop for circulation calculation and horizontal and vertical lines $A_2B_2$, $A_3B_3$ passing through core center for computing the average vortex core diameter, the plots of (b1) Y-velocity along line $A_2B_2$ and (b2) X-velocity along line $A_3B_3$ to estimate $d_{c-x}$ and $d_{c-y}$ respectively

4.6.2 Vortex Core Diameter ($D_c$)

The variation of core diameter with time for three different flow regimes in the study is presented in figure 11b. Two different core diameters: $d_{c-x}$ and $d_{c-y}$ along X and Y axes respectively are calculated by extracting the Y-velocity along the horizontal line $A_2B_2$ passing through the centre of the vortex core, as well as X-velocity along $A_3B_3$ as shown in figure 12 a, b. The distance between the location of two extremes of the velocity is considered as the core diameter. The average of $d_{c-x}$ and $d_{c-y}$ is taken as the vortex core diameter ($d_c$). Core diameter always increases with time during the formation of a vortex ring and starts to decrease after attaining a maximum value. The initial increase in $d_c$ with time is due to the continuous feeding of vorticity from the shear layer while the reason behind the subsequent decay in core size is viscous diffusion (see figure 6) as well as expelling of vorticity filament from the outer surface of the deformed vortex ring core (see figure 7) during propagation. Expelling of vorticity filament from the primary vortex ring surface is usually encountered during the interaction of the same with the vortices of opposite nature i.e. CRVRs for high PR cases. It is to be noted that the decline in core diameter for the case of PR = 50 is not due to pinching off; indeed, the strong CRVRs formed ahead of the primary start interacting with the latter, and it gets deformed with significant cancellation of vorticity. Furthermore, from the plot in figure 9b it is clear that as the PR (or $M_s$) increases the core diameter also increases.

For PR = 50, the CRVRs are strong enough to retain their distinct structure while encircling the primary ring (see figure 8f). Also, they are expelled toward the shear layer as observed in figure 8f which is the cause of the sudden drop in diameter in figure 11b for PR = 50. At this stage, the core is disintegrated and difficult to identify.

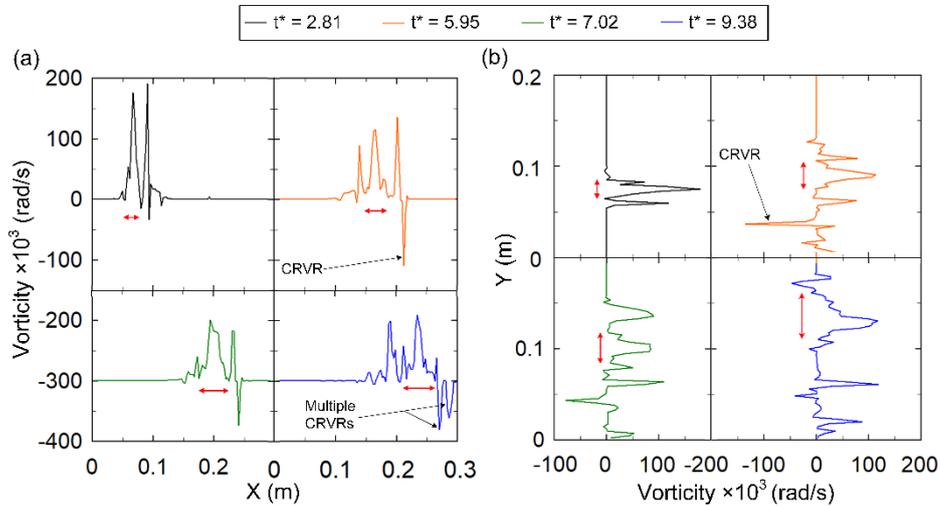

Figure 13. Z-Vorticity distribution through the centre of the core for the case of vortex ring evolution with PR = 50 air case at different times in the (a) horizontal direction (line $A_2B_2$ of figure 10) and (b) vertical direction (line $A_3B_3$ of figure 10)

In order to supplement the observed results of the vortex core evolution the vorticity distribution through the core centre in the horizontal and vertical directions for PR=50 case is plotted in figure 13. The distinct peak in the plot represents the centre of the vortex ring core, while another accompanying peak at initial time instant $t^* = 2.81$ designates the shear layer. It is observed that the height of such peaks decay with time, which is apparent as the primary ring begins to interact with CRVRs. Additionally, the multiple peaks at latter times signify vorticity expelling from the core, which ultimately approaches towards disintegration. The red arrow in the plot designates the qualitative scale of the core diameter, which continuously increases with time due to vorticity diffusion consistent to the observation in figure 9b.

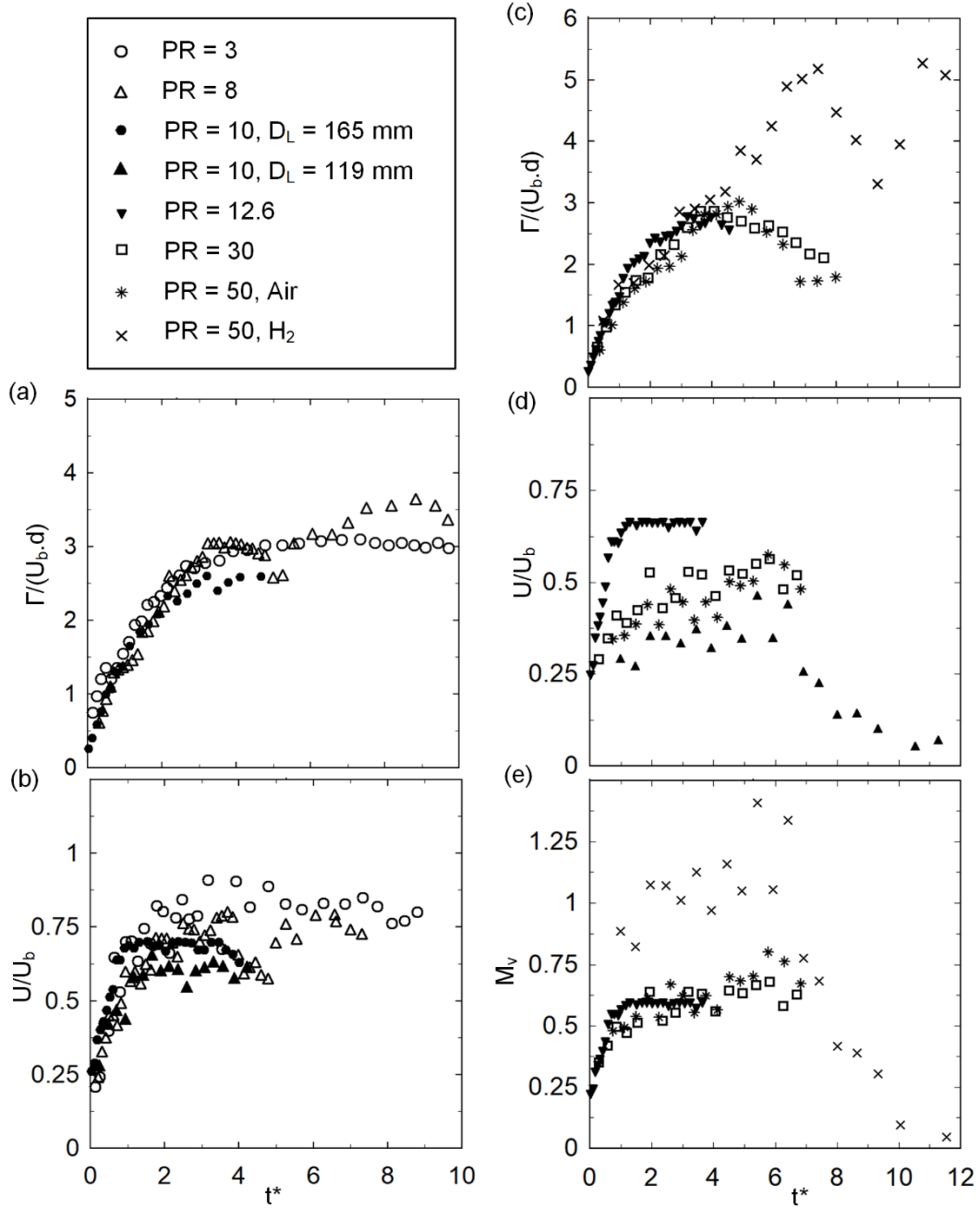

Figure 14. Evolution of compressible vortex ring characteristics with time. (a) Non-dimensional vortex ring circulation (b) Non-dimensional vortex ring translation velocity for lower PRs and (c), (d) the corresponding plots for higher PRs and (e) Variation of vortex ring Mach number ($M_v$)

4.6.3 Ring Circulation ($\Gamma$)

Circulation of a vortex ring is calculated by taking the line integral of the velocity vector along a closed-circuit enclosing the ring in the azimuthal plane as shown in figure 10a. Considering the variation of circulation with time in figure 14a and 14c, it is observed that circulation increases rapidly during the initial roll-up and growth of the ring. This is due to continuous feeding of vorticity into the core by the shear layer. After the vortex ring pinches off from the trailing jet, its circulation remains roughly constant during propagation as observed for PR =3 case which is again similar to incompressible flow as there is no induced shock or embedded shock present. The circulation after t* = 2.75 (t = 1328 μs) is observed to be almost steady for the duration of time the vortex ring is studied.

For PR = 8, circulation first attains a constant value then starts to decline at t* = 3.86 (t = 1050 μs) because, at this instant, primary vortex ring's interaction with a single tiny CRVR begins (figure 7 e, f). Soon, the effect of CRVR diminishes and the gradual increase in circulation of the vortex ring is observed after t* = 4.92 (t = 1338 μs, see figure 7 g, h) with some oscillation in its value due to the feeding of shear layer vortices into its core. The oscillation is circulation with time is linked with the frequency of merging of the discrete shear layer vortices into the primary core although apparently from visualization vortex rings appeared to be pinched-off.

In the higher PR cases (figure 14c) the primary vortex ring attains maximum circulation at a certain instant (t* = 4.86 or t = 650 μs for PR = 50, air), later the circulation decreases as the CRVRs are included in the calculations. Moreover, beyond this point of time (i.e. t* = 6.73, t = 900 μs for PR 50 case) circulation attains nearly constant value as the CRVRs encircle the primary ring before the core is severely stretched and nearly disintegrated.

Circulation for the case of PR = 50, $H_2$ is considerably more than the case of air at same PR, because of the higher strength of the shear layer and faster rate of vorticity feeding from the trailing jet to the vortex ring in the former. At t* = 7.06 (t = 434 μs, see figure 9e), the primary vortex is distinguishable from the opposite circulation slipstream and RVL. Beyond this time the circulation of the primary ring drops as it entrain some of the opposite circulation vortices into its core similar to air case. For air, this drop in circulation takes place after t* ~ 5 (see figure 8). The extent of the drop in primary ring circulation due to interaction with CRVRs is also the proof for the relatively higher strength of CRVRs in the case of hydrogen, which is consistent with the observations in numerical shadowgraphs (figure 8-9). Later, it is observed that the interaction of the same CRVRs with the primary makes the primary highly unstable and turbulent. The primary ring circulation in case of hydrogen increases again considerably. While this rise is not so significant in the case of the air; and the primary vortex ring is still compact as observed from the numerical shadowgraphs (figure 9). The exact reason for the sudden rise is difficult to ascertain during this complex interaction. However, the shrinking Mach disk at this time (see t* >10.54 in figure 9) indicates the weakening of the slipstream-CRVRs strength while the interaction of RVL Vortices-Primary and CRVRs results in the deformation of the vortex structure in such a way that the shear layer vorticity is suddenly gets engulfed into the primary ring (see figure 9, t* = 12.76 and greater). The inclusion of the shear-layer vorticity into the primary vortex ring results in continuous increase of circulation at this stage.

4.6.4 Ring Translation Velocity (U)

The effect of different PR's on the translation velocity (U) of the ring is shown in figure 12b and 12d. For PR = 3, the axial translation velocity increases gradually with time during the formation process and then attains a nearly uniform velocity of about $0.8U_b$. Unlike the low subsonic case (PR =3), a significant and interesting variation of translational velocity for PR = 8 is observed, which is due to the interaction of the primary vortex ring with a single CRVR and shear layer vortices (leapfrogging). The elliptic vortex core makes this interaction quite complex. After reaching maximum velocity at t* = 3.94 (t = 789 µs) (see figure 7e), the translation velocity decreases due to several effects: the single CRVR ahead, increases the diameter which is augmented by the leapfrogging of the shear layer vortex (see figure 7f) that results decrease in velocity. Note, that the shear layer vortex being stronger induces an opposing velocity to the core. The self-induced velocity also decreases due to the increase in diameter. As the CRVR encircles the primary vortex ring core (t* > 4.92 figure 7g) and diffuses, the diameter decreases (see figure 9a) and U increases. Subsequent increase in U is due to the merging of the shear layer vortex with the primary vortex ring and thus increasing its circulation as observed in figures 12a and 12c. The process continues as the shear layer vortices continue to feed into the primary vortex ring through a similar mechanism. A similar effect is not observed at higher PR cases (see figure 12d) as the slipstream CRVRs are stronger and multiple CRVRs are formed. At PR 12.6 U remains constant after initial acceleration as the effect of CRVRs and entrained shear layer vortices cancel each other's effects. In case of PR = 50 cases the translational velocity is highly influenced by the Biot-Savart induced velocity of several vortices and hence, remains nearly constant with a slow increase as shear layer vortices are fed into the core. Another important observation is at low PR cases the translational velocity attains nearly steady value of 0.7-0.8 $U_b$ (figure 14b) whereas, for high PR cases (figure 14d) the vortex ring velocity is much less (about $0.5U_b$). This is due to turbulent nature of the high PR vortex rings where multiple vortices form, breakdown into turbulent structure and thus loses considerable energy in turbulence production.

4.6.5 Vortex ring Mach number ($M_v$)

For the further characterization of vortex ring translation at high PR cases, the variation of vortex ring Mach number ($M_v$) with time is shown in figure 14e. It is evident that the change in the nature of expansions due to change in specific heat ratio, results in higher $U_b$ for PR =50, $H_2$ case and hence, $M_v$ become supersonic. Supersonic translational velocity of the vortex ring is not achieved in earlier studies in literature.

4.7 Jet Impulse at the Shock Tube Exit

Based on the results of the exit flow parameters of the jet (see figure 5), we compute the jet impulse, which has a major consequence on the formation of CRVRs (Dora et al. 2014). According to Dora et al. (2014), the peak-over-pressure magnitude and duration of the same are the parameters for CRVR formation. As per the findings of Ire et al. (2003), for transient jets, when pressure ratio continuously decreases, the Mach disc appears at the exit over-pressure ratio value less than 2.0 unlike that for steady jet as studied by Snedeker (1971). The exit over-pressure results in the impulse which until the disappearance of the Mach disc could be sufficient enough for creating a slipstream instability and ultimately CRVRs (Dora et al. 2014), Hereby, the method used by Dora et al. (2014) is adopted to calculate impulse: $t_{mnd}$ is the non-dimensionalized time of disappearance of the Mach disc. Impulse is non-dimensionalized as: $Im_{nd} = I/P_c t_c$. (Dora et al. 2014); where characteristics pressure $P_c = \rho_b U_b^2$, characteristics time $t_c = t/U_b d$. Thus calculated non-dimensionalized impulse for each case is tabulated in Table 3. According to Dora et al. (2014), the approximate value of the exit pressure at which the Mach disc disappearance occurs is $P/P_{atm} = 1.8$. It is known that the expanded flow at the shock tube exit is recompressed at the Mach disc position at a later time; the actual pressure signal felt just upstream of the Mach disc location at the time of its disappearance must have originated at the shock tube exit at an earlier time Dora et al. (2014). So, an impulse at the exit is calculated until the instant of time where $P/P_{atm} = 1.8$ is $I_{1.8}$ as well in a similar fashion and presented in Table 3.

Based on the analysis of impulse till the time of Mach disc disappearance, a plot of non-dimensionalized impulse versus $P_{max}/P_{atm}$ is constructed for each case of vortex ring evolution considered as shown in figure 5. It was observed by Dora et al. (2014) that there is no CRVR formation for the value of $Im_{nd} < 1.7$, while a single CRVR is formed for $1.7 < Im_{nd} < 2.6$ and multiple CRVRs are observed for $Im_{nd} > 2.6$. Our present study is also consistent with the earlier findings, as shown. It is inferred that, although a significant difference is observed in the vortex ring evolution and propagation process at high PR cases, the impulse-based criteria given by Dora et al. (2014) of formation of CRVR(s) holds the same.

Table 3. Details of Impulse and its consequence for different cases considered

| PR | $D_L$ | $U_b$ | $\dfrac{P_{max}}{P_{atm}}$ | $t_{mnd}$ | $I_m$ | $I_{mnd}$ | $t_{1.8nd}$ | $I_{1.8}$ | $I_{1.8nd}$ | MD | CRVR |
|---|---|---|---|---|---|---|---|---|---|---|---|
| 7 | 165 | 181 | 2.4 | 0 | 0 | 0 | 0.31 | 14 | 0.44 | No | No |
| 8 | 165 | 235 | 2.5 | 1.76 | 88 | 2.33 | 0.95 | 50 | 1.33 | Yes | Single |
| 10 | 119 | 280 | 2.8 | 2.09 | 105.2 | 2.47 | 1.86 | 82 | 1.9 | Yes | Single |
| 10 | 165 | 280 | 2.8 | 3.68 | 168 | 3.94 | 2.74 | 145 | 3.40 | Yes | Multi |
| 10 | 285 | 280 | 2.8 | 6.82 | 305 | 7.15 | 5.49 | 263 | 6.18 | Yes | Multi |
| 12.6 | 165 | 308 | 3.07 | 4.91 | 225 | 4.53 | 4.19 | 218 | 4.38 | Yes | Multi |
| 30 | 165 | 415 | 4.3 | 12.97 | 691 | 8.68 | 10.42 | 554 | 6.96 | Yes | Multi |
| 50 | 165 | 478 | 5.093 | 18.25 | 857 | 8.40 | 17.17 | 830 | 8.13 | Yes | Multi |
| 50 ($H_2$) | 165 | 1010 | 15.43 | 45.61 | 3238 | 28.80 | - | $> I_m$ | $> I_{mnd}$ | Yes | Multi |

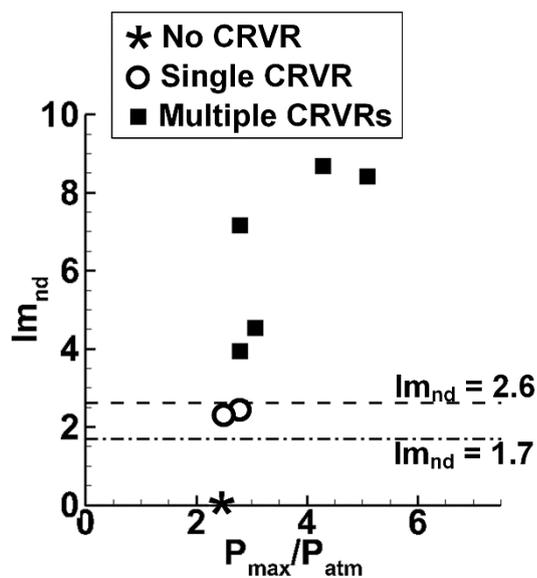

Figure 15. The non-dimensionalized impulse for various cases corresponding to $P_{max}/P_{atm}$

## 5. Conclusion

The compressible vortex rings generated at the open end of a shock tube has been investigated for various shock Mach numbers. The results indicate that the vortex ring forms at the exit of shock tube are free of shocks and the vortex ring remains compact and laminar during its evolution for lower PR (i.e. 3). For higher PRs, a vortex-induced shock forms in the vortex region along with an embedded shock underneath the vortex core. The primary vortex ring undergoes stretching and deformation due to supersonic expansion as well as the presence of induced shock and embedded shock. It is perceived that the vortex-induced shock and embedded shock grows stronger as the shock Mach number increases. For PR > 8, a shear layer is formed ahead of the vortex ring, which is very unstable and because of Kelvin-Helmholtz instability as multiple CRVRs are formed. The CRVRs form ahead of the primary vortex ring restricts the axial motion of the primary vortex ring. The ring characteristic parameters also get affected by this vortex-induced shock, embedded shock, and CRVRs. The results of the simulation with hydrogen as driver section gas shows that the flow is much more expanded than the case of air. Hydrogen at the driver section allowed to simulate a supersonic vortex ring where the additional flow features of reverse vortex layer and the tiny vortices are observed. Additionally, an impinging jet from incident shock is also observed along the axis in case of the supersonic vortex ring. The compressible vortex ring characteristics, including the ring diameter, translation velocity, circulation, and core size are computed and compared for the various cases studied herein. An excellent match of present results with the previously done experimental results is attained for varying $M_V$. This paper presents a comprehensive study of the compressible vortex ring and opens up the need for further investigation of even higher impulse, highest PR case considered here.

## 6. Acknowledgment

The authors would like to acknowledge the High-Performance Computing (HPC) Facility at IIT Kanpur (www.iitk.ac.in/cc) for providing the facility for computation and data analysis.

**Appendix**

**A1 Grid Convergence**

In this section, an analysis employing Richardson error estimator and Grid Convergence Index (GCI) (Choudhury et al. 2018) is reported. Estimation of Richardson error is performed by taking Grid-2 as the base grid. Solutions of (i) Z-vorticity ($\Omega_z$) and (ii) centreline velocity ($u_c$) are considered for two different assessments. The error values for fine and coarse grids are defined as:

$$E_1^{Fine} = \frac{\epsilon}{1-r^o} \quad (A1)$$

$$E_2^{Coarse} = \frac{r^o \epsilon}{1-r^o} \quad (A2)$$

Where, the error is calculated as: $\epsilon = f_2 - f_1$, with $f_1$ and $f_2$ being the solutions ($\Omega_z$ or $u_c$) for consequent grids. And the order of accuracy (o) is taken as 2.

Finally, Grid-Convergence Index (GCI) is computed with factor of safety ($F_S$) = 2 in order to achieve a uniform measure of grid independence which also takes care of the uncertainty in calculation of Richardson error estimator.

$$GCI = F_S |E| \quad (A3)$$

The table below shows the final calculated values of $E$ and $GCI$:

Table A1. Details of Richardson's error and Grid-Convergence Index for three different grids

| $r_{32}$ | $r_{21}$ | Variables | Position | $\epsilon_{21}$ (Fine) | $\epsilon_{32}$ (Fine) | $E_2^{Coarse}$ | $E_1^{Fine}$ | $GCI_{Coarse}$ | $GCI_{Fine}$ |
|---|---|---|---|---|---|---|---|---|---|
| 1.41 | 2.66 | $u_c$ | x = 0.04 m | -7 | 1 | 8.2 | -1.01 | 16.3 | 2.0 |
| | | | x = 0.19 m | 12 | -0.89 | -14.7 | 0.9 | 29.5 | 1.8 |
| | | $\Omega_z$ | y = 4.05E-2 m | 5151 | 100 | -5998.8 | -101.2 | 11997.6 | 202.4 |
| | | | y = 6.53E-2 m | -30504 | 3998 | 35524.7 | -4046.1 | 71049.5 | 8092.2 |

It is revealed that the error values for fine grid $E_1^{Fine}$ is reasonably lower than that for coarse grid $E_2^{Coarse}$. Finally, the optimum mesh size of ~0.7 million (Grid 2) is considered for all further simulations reported in the main manuscript.

**A2 Compressible Vortex Ring Evolution at PR = 12.6 and 30**

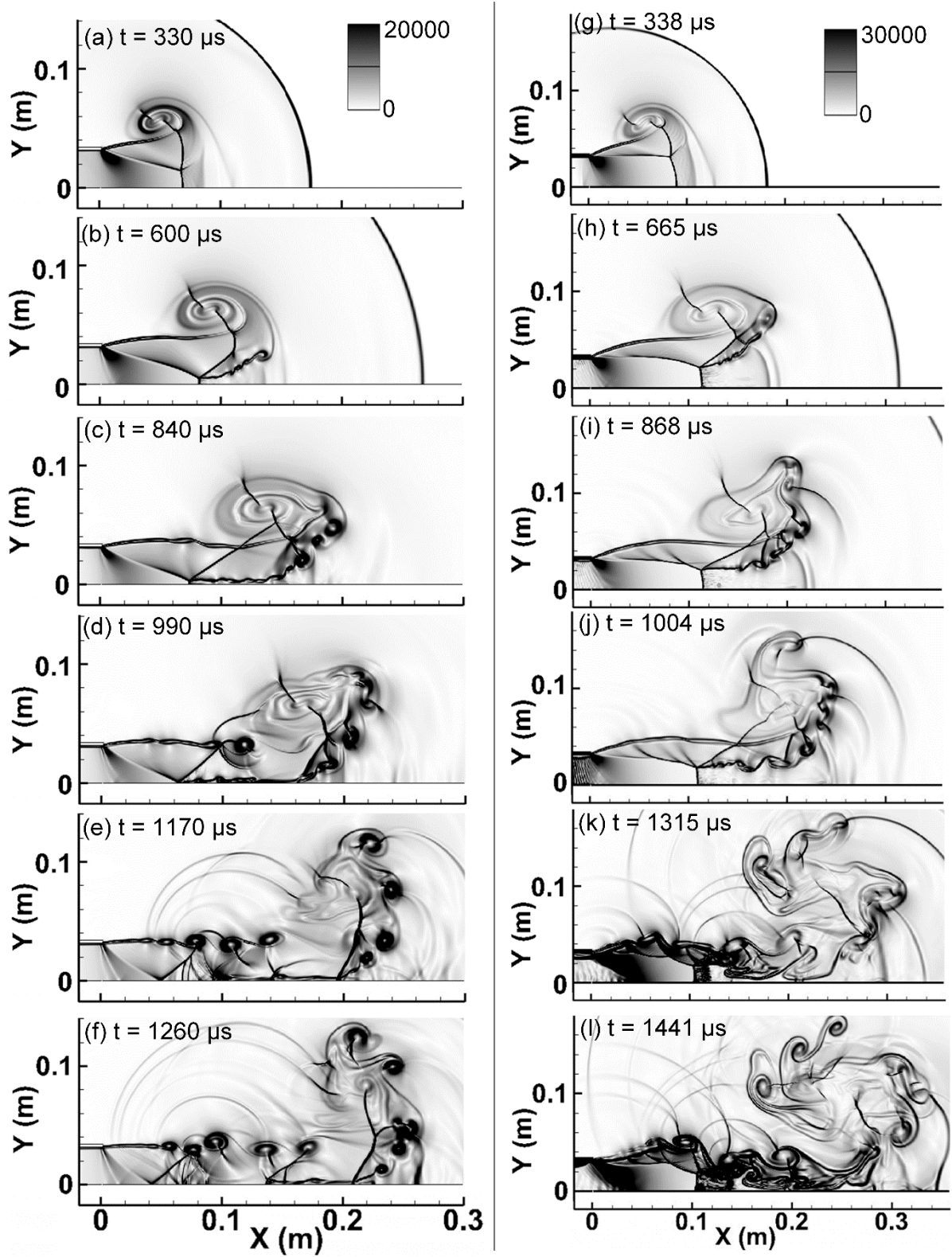

Figure A1. Numerical shadowgraphs of compressible vortex ring evolution with time for $D_L$ = 165 mm and PR = 12.6 (left column) and PR = 30 (right column).

Petersen, E. L., & Hanson, R. K. (2003). Improved turbulent boundary-layer model for shock tubes. *AIAA journal*, *41*(7), 1314-1322.

Poudel, S., Mishra, H., De, A., & Das, D. (2018) Numerical Investigation of Compressible Vortex Ring during Axial Interaction with Cylinder. In *Proceedings of the 7th International and 45th National Conference on Fluid Mechanics and Fluid Power (FMFP) December 10-12, 2018 (IIT Bombay, Mumbai, India)*

Riley, N., & Stevens, D. P. (1993). A note on leapfrogging vortex rings. *Fluid dynamics research*, *11*(5), 235.

Snedeker, R. S. (1971). A study of free jet impingement. Part 1. Mean properties of free and impinging jets. *Journal of fluid Mechanics*, *45*(2), 281-319.

Soni, R. K., & De, A. (2018). Investigation of mixing characteristics in strut injectors using modal decomposition. *Physics of Fluids*, *30*(1), 016108.

Soni, R. K., & De, A. (2018). Role of jet spacing and strut geometry on the formation of large scale structures and mixing characteristics. *Physics of Fluids*, *30*(5), 056103.

Soni, R. K., Arya, N., & De, A. (2019). Modal decomposition of turbulent supersonic cavity. *Shock Waves*, *29*(1), 135-151.

Sullivan, I. S., Niemela, J. J., Hershberger, R. E., Bolster, D., & Donnelly, R. J. (2008). Dynamics of thin vortex rings. *Journal of Fluid Mechanics*, *609*, 319-347.

Sun, M., & Takayama, K. (2003). A note on numerical simulation of vortical structures in shock diffraction. *Shock Waves*, *13*(1), 25-32.

Sun, M., & Takayama, K. (2003). Vorticity production in shock diffraction. *Journal of Fluid Mechanics*, *478*, 237-256.

Takayama, F., Ishii, Y., Sakurai, A., & Kambe, T. (1993). Self-intensification in shock wave and vortex interaction. *Fluid dynamics research*, *12*(6), 343.

Tokugawa, N., Ishii, Y., Sugano, K., Takayama, F., & Kambe, T. (1997). Observation and analysis of scattering interaction between a shock wave and a vortex ring. *Fluid dynamics research*, *21*(3), 185.